\documentstyle[12pt]{article}

\makeatletter
\@addtoreset{equation}{section}
\makeatother

\let\la=\label  
   
 \def\bd{\begin{document}} \def\ed{\end{document}}
\def\ds{\documentstyle} \let\fr=\frac \let\bl=\bigl \let\br=\bigr
\let\Br=\Bigr \let\Bl=\Bigl 
\let\bm=\bibitem
\let\na=\nabla
\let\pa=\partial \let\ov=\overline 
\newcommand{\be}{\begin{equation}} 
\newcommand{\ee}{\end{equation}} 
\def\ba{\begin{array}}
\def\ea{\end{array}}
\newcommand{\ho}[1]{$\, ^{#1}$}
\newcommand{\hoch}[1]{$\, ^{#1}$}
\newcommand{\bea}{\begin{eqnarray}} 
\newcommand{\eea}{\end{eqnarray}} 
\newcommand{\ra}{\rightarrow}
\newcommand{\lra}{\longrightarrow}
\newcommand{\Lra}{\Leftrightarrow}
\newcommand{\ap}{\alpha^\prime}
\newcommand{\bp}{\tilde \beta^\prime}
\newcommand{\tr}{{\rm tr} }
\newcommand{\Tr}{{\rm Tr} } 
\newcommand{\NP}{Nucl. Phys. }
\newcommand{\tamphys}{\it Center for Theoretical Physics\\
Texas A\&M University, College Station, Texas 77843}

\newcommand{\auth}{M. J. Duff\footnote{Research supported in  part by 
NSF Grant PHY-9411543.}}

\begin{document}

\hfill{CTP-TAMU-33/96}

\hfill{hep-th/9608117}

\vspace{24pt}

\begin{center}
{ \large 
{\bf ~~~~~~~~~~~~~~~~~~~~M-THEORY 
\newline
(THE THEORY FORMERLY KNOWN AS STRINGS)\footnote{Based on talks given at the
Geometry and Physics Conference, Warwick, U. K., March 1996; the SUSY 96
conference, Maryland, U.S.A., June 1996; the CERN Duality Workshop, Geneva,
Switzerland, June 1996; and the International School of Subnuclear Physics, 
Erice, Italy, July 1996.}}}

\vspace{36pt}

\auth

\vspace{10pt}

{\tamphys}

\vspace{44pt}

\underline{ABSTRACT}

\end{center}

Superunification underwent a major paradigm shift in 1984 when 
eleven-dimensional supergravity was knocked off its pedestal by
ten-dimensional superstrings. This last year has witnessed a new shift of
equal proportions: perturbative ten-dimensional superstrings have in their
turn been superseded by a new non-perturbative theory called {\it 
$M$-theory}, which describes supermembranes and superfivebranes, which
subsumes all five consistent string theories and whose low energy limit
is, ironically, eleven-dimensional supergravity. In particular,
six-dimensional  string/string duality follows from membrane/fivebrane
duality by compactifying $M$-theory on  $ S^1/Z_2 \times K3$
(heterotic/heterotic duality) or $ S^1 \times K3$ (Type $IIA$/heterotic
duality) or $S^1/Z_2 \times T^4$ (heterotic/Type $IIA$ duality)  or $S^1
\times T^4$ (Type $IIA$/Type $IIA$ duality).

{\vfill\leftline{}\vfill}

\pagebreak
\setcounter{page}{1}

\section{Ten to eleven: it is not too late}
\la{Introduction}

The maximum spacetime dimension in which one can formulate a consisistent
supersymmetric theory is eleven\footnote{The field-theoretic reason is
based on the prejudice that there be no massless particles with spins
greater than two \cite{Nahm}. However, as discussed in section
(\ref{twelve}), $D=11$ emerges naturally as the maximum dimension admitting
super $p$-branes in Minkowski signature.}.  For this reason in the
early  1980's many physicists looked to $D=11$ supergravity \cite{Julia},
in the hope that it might provide that superunification \cite{Pope} they
were all looking for. Then in 1984 superunification underwent a major
paradigm shift: eleven-dimensional supergravity was knocked off its
pedestal by ten-dimensional superstrings \cite{Green}, and eleven
dimensions fell out of favor. This last year, however, has witnessed a new
shift of  equal proportions: perturbative ten-dimensional superstrings
have in their turn been superseded by a new non-perturbative theory called
{\it  $M$-theory}, which describes (amongst other things) supersymmetric
extended objects with two spatial dimensions ({\it supermembranes}), and
five spatial dimensions ({\it superfivebranes}), which subsumes all five
consistent string theories and whose low energy limit is, ironically,
eleven-dimensional supergravity.

The reason for this reversal of fortune of eleven dimensions is due, in 
large part, to the 1995 paper by Witten \cite{Wittenv}. One of the
biggest problems  with $D=10$ string theory \cite{Green} is that there are
{\it five} consistent string theories:  Type $I$ $SO(32)$, heterotic 
$SO(32)$, heterotic $E_8 \times E_8$,  Type $IIA$ and Type $IIB$. As a
candidate for a unique {\it theory of everything}, this is clearly an
embarrassment of riches.  Witten put forward a convincing case that 
this distinction is just an artifact of perturbation theory and that
non-perturbatively these five theories are, in fact, just different
corners of a deeper theory. Moreover, this deeper theory, subsequently
dubbed {\it $M$-theory}, has $D=11$ supergravity as its low energy limit!
Thus the five string theories and $D=11$ supergravity represent six
different special points\footnote{Some authors take the phrase {\it
$M$-theory} to refer merely to this sixth corner of the moduli space. With
this definition, of course, $M$-theory is no more fundamental than the
other five corners. For us, {\it $M$-theory} means the whole kit and
caboodle.} in the moduli space of $M$-theory.  The small parameters of
perturbative string theory are provided by $<e^{\Phi}>$, where $\Phi$ is
the dilaton field, and $<e^{{\sigma}_i}>$ where ${\sigma}_i$ are the
moduli fields which arise after compactification. What makes $M$-theory at
once intriguing and yet difficult to analyse is that in $D=11$ there is
neither dilaton nor moduli and hence the theory is intrinsically
non-perturbative. Consequently, the ultimate meaning of $M$-theory is
still unclear, and Witten has suggested that in the meantime, $M$ should
stand for ``Magic'', ``Mystery'' or ``Membrane'', according to taste.

The relation between the membrane and the fivebrane in $D=11$ is analogous 
to the relation between electric and magnetic charges in $D=4$.  In fact
this is more than an analogy: electric/magnetic duality in $D=4$
string theory \cite{Font,Rey} follows as a consequence of string/string
duality in $D=6$ \cite{Duffstrong}. The main purpose of the present paper
is to show  how $D=6$ string/string duality
\cite{Luloop,Khurifour,Lublack,Minasian1,Hulltownsend,Khuristring,Wittenv} 
follows, in its turn, as a consequence of membrane/fivebrane duality in
$D=11$. In particular, heterotic/heterotic duality, Type
$IIA$/heterotic duality, heterotic/Type $IIA$ duality
and Type $IIA$/Type $IIA$ duality follow from membrane/fivebrane duality
by compactifying $M$-theory on $ S^1/Z_2 \times K3$ \cite{DMW}, $ S^1
\times K3$ \cite{Minasian2}, $S^1/Z_2 \times T^4$  and $S^1 \times T^4$,
respectively.

First, however, I want to pose the question:``Should we have been surprised
by the eleven-dimensional origin of string theory?''
 
\section{Type II A\&M theory}
\label{cool}

The importance of eleven dimensions is no doubt surprising from the point 
of view of perturbative string theory; from the point of view of membrane
theory, however,  there were already tantalizing hints in this direction:

($i$) {\bf K3 compactification}

The four-dimensional compact manifold $K3$ plays a ubiquitous role in much 
of present day $M$-theory.  It was first introduced as a compactifying
manifold in 1983 \cite{Nilsson2} when it was realised that the number of
unbroken supersymmetries surviving compactification in a Kaluza-Klein
theory depends on the {\it holonomy} group of the extra dimensions. By
virtue of its $SU(2)$ holonomy, $K3$ preserves precisely half of the
supersymmetry.  This means, in particular, that an $N=2$ theory on $K3$
has the same number of supersymmetries as an $N=1$ theory on $T^4$, a
result which was subsequently to prove of vital importance for
string/string duality.  In 1986, it was pointed out \cite{Nilsson1} that
$D=11$ supergravity on $R^{10-n}\times K3\times T^{n-3}$  \cite{Nilsson2}
and the $D=10$ heterotic string on $R^{10-n}\times  T^{n}$ \cite{Narain}
not only have the same supersymmetry but also the same moduli spaces of
vacua, namely     
\begin{equation}   
{\cal M}=\frac{SO(16+n,n)}{SO(16+n)
\times SO(n)} 
\label{moduli}  
\end{equation} 
It took almost a decade for this ``coincidence'' to be explained but we now 
know that
$M$-theory on $R^{10-n}\times K3\times T^{n-3}$ is dual to the heterotic 
string on $R^{10-n}\times T^{n}$.

($ii$) {\bf Superstrings in D=10 from supermembranes in D=11}

Eleven dimensions received a big shot in the arm in 1987 when the $D=11$ 
supermembrane 
was discovered \cite{Bergshoeff2}.   The bosonic sector of its $d=3$
worldvolume Green-Schwarz action is given by: 
\begin{eqnarray}
S_3&=&T_3\int d^3\xi\biggl[-{1\over2}\sqrt{-\gamma}\gamma^{ij}
\partial_i X^M\partial_j X^N G_{MN}(X) +{1\over2}\sqrt{-\gamma}\nonumber\\
&&\qquad\qquad
-{1\over3!}\epsilon^{ijk}\partial_i X^M\partial_j X^N\partial_k X^P
C_{MNP}(X)\biggr]
\label{membrane}
\end{eqnarray}
where $T_3$ is the membrane tension, $\xi^i$ ($i=1,2,3$) are the
worldvolume coordinates, $\gamma^{ij}$ is the worldvolume metric and
$X^M(\xi)$ are the spacetime coordinates $(M=0,1,\ldots,10)$.  Kappa
symmetry \cite{Bergshoeff2} then demands that the background metric
$G_{MN}$ and background 3-form potential $C_{MNP}$ obey the classical
field equations of $D=11$ supergravity \cite{Julia}, whose bosonic action 
is
\begin{equation}
I_{11}=\frac{1}{2\kappa_{11}^2}\int d^{11}x\sqrt{-G}
\left[R_G-\frac{1}{2\cdot4!}K_4{}^2\right]
-\frac{1}{12\kappa_{11}^2} \int C_3\wedge K_4 \wedge K_4 
\label{supergravity11}
\end{equation}
where $K_4=dC_3$ is the 4-form field strength. In particular, $K_4$
obeys the field equation
\begin{equation}
d*K_4=-{1\over2}K_4{}^2
\label{equation4}
\end{equation}
and the Bianchi identity
\begin{equation}
dK_4=0
\label{Bianchi4}
\end{equation}

It was then pointed out \cite{Howe} that in an $R^{10} \times S^1$ topology
the weakly coupled $(d=2,D=10)$  Type $IIA$ superstring follows by
wrapping the $(d=3,D=11)$ supermembrane around the circle in the limit
that its radius $R$ shrinks to zero. In particular, the Green-Schwarz
action of the string follows in this way from the Green-Schwarz action of
the membrane. It was necessary to take this $R \rightarrow 0$ limit in
order to send to infinity the masses of the (at the time) unwanted
Kaluza-Klein modes which had no place in weakly coupled Type $IIA$ theory.
The $D=10$ dilaton, which governs the strength of the string coupling, is
just a component of the $D=11$ metric. 

A critique of superstring orthodoxy {\it circa} 1987, and its failure to
accommodate the eleven-dimensional supermembrane, may be found in
\cite{Duffnot}.

($iii$) {\bf U-duality (when it was still non-U)}

Based on considerations of this $D=11$ supermembrane, which on further 
compactification treats the dilaton and  moduli fields on the same footing,
it was conjectured \cite{Luduality} in 1990 that discrete subgroups  of
all the old non-compact global symmetries of compactified supergravity
\cite{Scherk,Cremmer} (e.g $SL(2,R)$, $O(6,6)$, $E_7$) should be
promoted to duality symmetries of the supermembrane.  Via the above
wrapping around $S^1$, therefore, they should be also be inherited by the 
Type $IIA$ string \cite{Luduality}.

($iv$) {\bf D=11 membrane/fivebrane duality}

In 1991, the supermembrane was recovered as an elementary solution of 
$D=11$ supergravity which preserves half of the spacetime supersymmetry
\cite{Stelle}. Making the three/eight split $X^M=(x^{\mu},y^m)$ where
$\mu=0,1,2$ and $m=3,...,10$, the metric is given by 
\be
ds^2=(1+k_3/y^6)^{-2/3}dx^{\mu}dx_{\mu}+
(1+k_3/y^6)^{1/3}(dy^2+y^2d\Omega_7{}^2)
\ee
and the four-form field strength by
\be
\tilde K_7 \equiv *K_4=6k_3\epsilon_7
\ee
where the constant $k_3$ is given by
\be
k_3=\frac{2\kappa_{11}{}^2T_3}{\Omega_7}
\ee
Here $\epsilon_7$ is the volume form on $S^7$ and $\Omega_7$ is the volume. 
The mass 
per unit area of the membrane ${\cal M}_3$ is equal to its tension: 
\be
{\cal M}_3=T_3
\ee
This {\it elementary} solution is a singular solution of the supergravity 
equations coupled to a supermembrane source and carries a Noether
``electric'' charge  
\be
Q=\frac{1}{\sqrt{2}\kappa_{11}}\int_{S^7}(*K_4 + C_3 \wedge K_4)
=\sqrt{2}\kappa_{11}T_3 
\ee
Hence the solution saturates the Bogomol'nyi bound
$\sqrt{2}\kappa_{11}{\cal M}_3\geq Q$. This is a consequence of the
preservation of half the supersymmetries  which is also intimately linked
with the worldvolume kappa symmetry. The zero modes of this solution
belong to a $(d=3,n=8)$ supermultiplet consisting of eight scalars and
eight spinors $(\phi^I,\chi^I)$, with $I=1,...,8$, which correspond to the
eight Goldstone bosons and their superpartners associated with breaking of
the eight translations transverse to the membrane worldvolume.  

In 1992, the superfivebrane was discovered as a soliton solution of $D=11$ 
supergravity also preserving half the spacetime supersymmetry
\cite{Gueven}. Making the six/five split $X^M=(x^{\mu},y^m)$ where
$\mu=0,1,2,3,4,5$ and $m=6,...,10$, the metric is given by
\be
ds^2=(1+k_6/y^3)^{-1/3}dx^{\mu}dx_{\mu}+
(1+k_6/y^3)^{2/3}(dy^2+y^2d\Omega_4{}^2)
\ee
and the four-index field-strength by
\be
K_4=3k_6\epsilon_4
\ee
where the fivebrane tension ${\tilde T}_6$ is related to the constant $k_6$
by
\be
k_6=\frac{2\kappa_{11}{}^2{\tilde T}_6}{3\Omega_4}
\ee
Here $\epsilon_4$ is the volume form on $S^4$ and $\Omega_4$ is the volume.
The mass per unit $5$-volume of the fivebrane ${\cal M}_6$ is equal to its 
tension:
\be
{\cal M}_6={\tilde T}_6
\ee
This {\it solitonic} solution is a non-singular solution of the source-free
equations 
and carries a topological ``magnetic'' charge   
\be 
P=\frac{1}{\sqrt{2}\kappa_{11}}\int_{S^4}K_4=\sqrt{2}\kappa_{11}{\tilde
T}_6 
\ee
Hence the solution saturates the Bogomol'nyi
bound $\sqrt{2}\kappa_{11}{\cal M}_6\geq P$. Once again, this is a
consequence of the preservation of half the  supersymmetries. The
covariant action for this $D=11$ superfivebrane is still unknown (see
\cite{Ortin,Sezgin} for recent progress) but consideration  of the soliton zero
modes \cite{Gibbonstownsend,Khuristring,Townsendeleven} means that the
gauged fixed action must be described by the same chiral antisymmetric
tensor multiplet $(B^-{}_{\mu\nu},\lambda^I,\phi^{[IJ]})$ as that of the
Type $IIA$ fivebrane \cite{Callan1,Callan2}. Note that in addition to the
five scalars corresponding to the five translational Goldstone bosons, there
is also a $2$-form $B^-{}_{\mu\nu}$ whose $3$-form field strength is
anti-self dual and which describes three degrees of freedom.    

The electric and magnetic charges obey a Dirac quantization rule
\cite{Nepomechie,Teitelboim} 
\be
QP=2\pi n \qquad n={\rm integer}
\ee
Or, in terms of the tensions \cite{Luelem,Lublack},
\begin{equation}
2\kappa_{11}{}^2 T_3 {\tilde T}_6 =2\pi  n
\label{Dirac11}
\end{equation}
This naturally suggests a $D=11$ membrane/fivebrane duality. Note that
this reduces the three dimensionful parameters $T_3$, ${\tilde T}_6$ and
$\kappa_{11}$ to two.   Moreover, it was recently shown \cite{Minasian2}
that they are not independent.  To see this, we note from (2.2)
that $C_3$ has period $2\pi/T_3$ so that $K_4$ is quantized according to  
\begin{equation} \int K_4={2\pi n\over T_3}\,\,\,\,\,n=integer \label{kquant}
\end{equation}
Consistency of such $C_3$ periods with the spacetime action,
(\ref{supergravity11}), gives the relation\footnote{This 
corrects a factor of two error in \cite{Minasian2} and brings us into
agreement with a subsequent $D$-brane derivation 
\cite{Schwarzpower} of (\ref{tension}). I am grateful to Shanta De Alwis
\cite{Dealwis} for pointing out the source of the error.}  
\begin{equation}
{(2\pi)^2\over\kappa_{11}{}^2T_3^3}\in 2Z
\label{eq:k11t3}
\end{equation}
From (\ref{Dirac11}), this may also be written as 
\begin{equation}
2\pi {\tilde T_6\over T_3{}^2}\in Z
\label{eq:newdirac}
\end{equation}
Thus the tension of the singly charged fivebrane is given by
\be
\tilde T_6=\frac{1}{2\pi}T_3{}^2
\la{tension}
\ee
 
($v$) {\bf Hidden eleventh dimension}

We have seen how the $D=10$ Type $IIA$ string follows from $D=11$.  Is it 
possible to go the other way and discover an eleventh dimension hiding in
$D=10$? In 1993, it was recognized \cite{Luscan} that by dualizing a
vector into a scalar on the gauge-fixed  $d=3$ worldvolume of the Type
$IIA$ supermembrane, one increases the number of worldvolume scalars (i.e
transverse dimensions) from $7$ to $8$ and hence obtains the corresponding
worldvolume action of the $D=11$ supermembrane.  Thus the $D=10$ Type
$IIA$ theory contains a hidden $D=11$ Lorentz invariance! This device was
subseqently used \cite{TownsendM,Schmidhuber} to demonstrate the
equivalence of the actions of the $D=10$ Type $IIA$ membrane and the
Dirichlet twobrane \cite{Polchinski}.

($vi$) {\bf U-duality}

Of the conjectured Cremmer-Julia symmetries referred to in $(iii)$ above, 
the case for a target space $O(6,6;Z)$ ({\it $T$-duality}) in perturbative
string theory had already been made, of course \cite{Giveonreview}. 
Stronger evidence for an $SL(2,Z)$ ({\it $S$-duality}) in string theory
was subsequently provided in \cite{Font,Rey} where it was pointed out that
it corresponds to a {\it non-perturbative} electric/magnetic symmetry.

In 1994, stronger evidence for the combination of $S$ and $T$ into a 
discrete duality of Type $II$ strings, such as $E_7(Z)$ in $D=4$, was
provided in \cite{Hulltownsend}, where it was dubbed {\it $U$-duality}. 
Moreover, the BPS spectrum necessary for this $U$-duality was given an
explanation in terms of the wrapping of either the $D=11$ membrane or
$D=11$ fivebrane around the extra dimensions. This 
paper also conjectured a non-perturbative $SL(2,Z)$ of the
Type $IIB$ string in $D=10$. 

($vii$) {\bf Black Holes}

In 1995, it was conjectured \cite{Townsendeleven} that the $D=10$ Type 
$IIA$ superstring should be identified with the $D=11$ supermembrane
compactified on $S^1$, even for large $R$. The $D=11$ Kaluza-Klein modes
(which, as discussed in $(ii)$ above, had no place in the {\it
perturbative} Type $IIA$ theory) were interpreted as charged extreme black
holes of the Type $IIA$ theory.

($viii$) {\bf D=11 membrane/fivebrane duality and anomalies}

Membrane/fivebrane duality interchanges the roles of field equations and 
Bianchi
identities. From (\ref{equation4}), the fivebrane Bianchi identity reads  
\begin{equation}
d\tilde K_7=-{1\over2}K_4{}^2
\end{equation}
However, it was recognized in 1995 that such a Bianchi identity will in
general require gravitational Chern-Simons corrections arising from a
sigma-model anomaly on the fivebrane worldvolume \cite{Minasian2}
\begin{equation}
d\tilde K_7=-{1\over2}K_4^2 + {2\pi\over {\tilde T}_6}{\tilde X}_8
\label{Bianchi7}
\end{equation}
where the $8$-form polynomial
${\tilde X}_8$, quartic in the gravitational curvature $R$, 
describes the Lorentz $d=6$ worldvolume anomaly  of the $D=11$
fivebrane.  Although the covariant fivebrane action is unknown, we know
that the gauge fixed theory is
described by the chiral antisymmetric tensor multiplet $(B^-_{\mu\nu},
\lambda^I, \phi^{[IJ]})$, and it is a straightforward matter to read
off the anomaly polynomial from the literature.  See, for example
\cite{Alvarez}.  We find 
\begin{equation}
{\tilde X}_8={1\over(2\pi)^4}[-{1\over768}(\tr R^2)^2+{1\over192}\tr R^4]
\la{X}
\end{equation}
Thus membrane/fivebrane duality predicts a spacetime correction to the
$D=11$ supermembrane action \cite{Minasian2}
\begin{equation}
I_{11}(Lorentz)=T_3\int C_3\wedge
{1\over(2\pi)^4}[-{1\over768}(\tr R^2)^2+{1\over192}\tr R^4] 
\end{equation}
Such a correction was also derived in a somewhat different way in
\cite{Wittenfive}.  This prediction is intrinsically
$M$-theoretic, with no counterpart in ordinary $D=11$
supergravity. However, by simultaneous dimensional reduction \cite{Howe}
of $(d=3,D=11)$ to $(d=2,D=10)$ on $S^1$, it translates into
a corresponding prediction for the Type $IIA$ string:  
\begin{equation}
I_{10}(Lorentz)=T_2\int B_2\wedge
{1\over(2\pi)^4}[-{1\over768}(\tr R^2)^2+{1\over192}\tr R^4] 
\end{equation}
where $B_2$ is the string $2$-form and $T_2=1/2\pi \alpha'$ is the string
tension.

As a consistency check we can compare this prediction with 
previous results found by explicit string one-loop calculations. These
have been done in two ways: either by computing directly in $D=10$ the
Type $IIA$ anomaly polynomial \cite{Liu} following
\cite{Schellekens},
or by compactifying to $D=2$ on an $8$-manifold $M$ and computing the
$B_2$ one-point function \cite{Vafawitten2}.  We indeed find agreement. 
Thus using $D=11$ membrane/fivebrane duality we have correctly
reproduced the corrections to the $B_2$ field equations of the $D=10$
Type $IIA$ string (a mixture of tree-level and string one-loop effects)
starting from the Chern-Simons corrections to the Bianchi identities of
the $D=11$ superfivebrane (a purely tree-level effect). It would be 
interesting to know, on the membrane side, what calculation in $D=11$
$M$-theory, when reduced on  $S^1$, corresponds to this one-loop Type
$IIA$ string amplitude calculation in $D=10$. Understanding this may well
throw a good deal of light on the mystery of what $M$-theory really is!

($ix$) {\bf Heterotic string from fivebrane wrapped around $K3$}

In 1995 it was shown that, when wrapped around $K3$ with its $19$ self-dual and $3$
anti-self-dual $2$-forms, the $d=6$ worldvolume fields of the $D=11$
fivebrane (or Type $IIA$ fivebrane)
$(B^-{}_{\mu\nu},\lambda^I,\phi^{[IJ]})$ reduce to   the $d=2$ worldsheet
fields of the heterotic string in $D=7$ (or $D=6$)
\cite{Townsendseven,Harveystrominger}. The $2$-form yields $(19,3)$ left 
and right moving bosons, the spinors yield $(0,8)$ fermions and the
scalars yield $(5,5)$ which add up to the correct worldsheet degrees of
freedom of the heterotic string \cite{Townsendseven,Harveystrominger}.

A consistency check is provided \cite{Minasian2} by the derivation of the
Yang-Mills and Lorentz Chern-Simons corrections to the Bianchi identity of
the heterotic string starting from the fivebrane Bianchi identity given
in $(viii)$ . Making the seven/four split $X^M=(x^{\mu},y^m)$ where
$\mu=0,...,6$ and $m=7,8,9,10$, the original set of $D=11$ fields
may be decomposed in a basis of harmonic $p$-forms on $K3$.  In
particular, we expand $C_3$ as 
\begin{equation}
C_3(X)=C_3(x)+{1\over2T_3}\sum C_1^I(x)\omega_2^I(y)
\end{equation}
where $\omega_2^I$, $I=1,\ldots,22$ are an integral basis of $b_2$ harmonic
two-forms on $K3$.  Following \cite{Minasian1}, let us define the dual
string $3$-form ${\tilde H}_3$ by
\begin{equation}
T_2{\tilde H}_3
={\tilde T}_6\int_{K3}{\tilde K}_7,
\end{equation}
The dual string Lorentz anomaly polynomial, ${\tilde X}_4$, is given by
\begin{eqnarray}
{\tilde X}_4=\int_{K3}\tilde X_8&=
&{1\over(2\pi)^2}{1\over192}\tr R^2 p_1(K3)\nonumber\\&
\end{eqnarray}
where $p_1(K3)$ is the Pontryagin number of $K3$
\begin{equation}
p_1(K3)=-{1\over8\pi^2}\int_{K3}\tr R_0^2=-48
\end{equation}
We may now integrate (\ref{Bianchi7}) over $K3$, using (\ref{tension})
to find   
\begin{equation}
d\tilde H_3=-{\alpha'\over4}[K_2^IK_2^Jd_{IJ}+\tr R^2]
\label{eq:d7bi}
\end{equation}
where $K_2^I=dC_1^I$ and where $d_{IJ}$ is the intersection matrix on $K3$,
given by 
\begin{equation}
d_{IJ} = \int_{K3}\omega_2^I\wedge\omega_2^J
\end{equation}
which has $b_2^+=3$ positive and $b_2^-=19$ negative eigenvalues.  Thus
we see that this form of the Bianchi identity corresponds to a $D=7$
toroidal compactification of a heterotic string at a generic point on
the Narain lattice \cite{Narain}.  Thus we have reproduced the $D=7$
Bianchi identity of the heterotic string, starting from the $D=11$
fivebrane.

For use in section (\ref{duality}), we note that if we replace $K3$ by
$T^4$ in the above derivation, the $2$-form now yields $(3,3)$ left  and
right moving bosons, the spinors now yield $(8,8)$ fermions and the scalars
again yield $(5,5)$ which add up to the correct worldsheet degrees of
freedom of the Type $IIA$ string. In this case, the Bianchi identity
becomes $d{\tilde H}_3=0$ as it should be. 

($x$) {\bf N=1 in D=4}

Also in 1995 it was noted
\cite{Cadavid,Papadopoulos,Schwarz3,Lowe,Chaudhuri,Acharya,Aspinwall4}
that $N=1$ heterotic strings can be dual to $D=11$ supergravity
compactified on seven-dimensional spaces of $G_2$ holonomy which also
yield $N=1$ in $D=4$ \cite{Pope2}. 

($xi$) {\bf Non-perturbative effects}

Also in 1995 it was shown \cite{Becker} that membranes and fivebranes of
the Type $IIA$ theory, obtained by compactification on $S^1$, yield
$e^{-1/g_s}$ effects, where $g_s$ is the string coupling.

($xii$) {\bf SL(2,Z)}

Also in 1995, strong evidence was provided for identifying the Type $IIB$
string on  $R^9 \times S^1$ with $M$-theory on $R^9 \times T^2$
\cite{Schwarzpower,Aspinwall4}.   In particular, the conjectured $SL(2,Z)$
of the Type $IIB$ theory discussed in $(vi)$ above is just the modular
group of the $M$-theory torus\footnote{Two alternative explanations of this
$SL(2,Z)$ had previously been given: (a) identifying it with the
$S$-duality \cite{Minasian2} of the  $d=4$ Born-Infeld worldvolume theory
of the self-dual Type $IIB$ superthreebrane \cite{Luthree}, and (b)
using the four-dimensional heterotic/Type $IIA$/Type $IIB$ triality
\cite{Duffliurahmfeld} by noting that this $SL(2,Z)$, while
non-perturbative for the Type $IIB$ string, is perturbative for the
heterotic string.}

($xiii$) {\bf ${\rm E_8 \times E_8}$ heterotic string}

Also in 1995 (that {\it annus mirabilis!}), strong evidence was provided
for identifying the  $E_8 \times E_8$  heterotic string \footnote{It is
ironic that, having hammered the final nail in the coffin of $D=11$
supergravity by telling us that it can never yield a {\it chiral} theory
when compactified on a manifold \cite{Wittenchiral}, Witten pulls it out
again by telling us that it does yield a chiral theory when compactified on
something that is not a manifold!} on $R^{10}$ with $M$-theory on $R^{10}
\times S^1/Z_2$ \cite{Horavawitten}.

This completes our summary of $M$-theory before $M$-theory was cool.  The
{\it phrase} $M$-theory (though, as I hope to have shown, not the  {\it
physics} of $M$-theory) first made its appearance in October 1995
\cite{Schwarzpower,Horavawitten}. This was also the month that it was
proposed \cite{Polchinski} that the Type $II$ $p$-branes carrying
Ramond-Ramond charges can be given an exact conformal field theory
description via open strings with Dirichlet boundary conditions, thus
heralding the era of $D-branes$. Since then, evidence in favor of
$M$-theory and $D$-branes has been appearing daily on the internet, 
including applications to black holes \cite{Stromvafa}, length scales
shorter than the string scale \cite{Shenker} and even phenomenology
\cite{Dine,Petr}.  We refer the reader to the review by Schwarz
\cite{Schwarzreview} for these more recent developments in $M$-theory, to
the  review by Polchinski \cite{Polchinski2} for developments in
$D$-branes and to the paper by Aharony, Sonnenschein and Yankielowicz
\cite{Aharony} for the connection between the two (since $D$-branes are
intrinsically ten-dimensional and $M$-theory is eleven-dimensional, this
is not at all obvious).  Here, we wish to focus on a specific application
of $M$-theory, namely the derivation of string/string dualities.

\section{String/string duality from M-theory}
\label{duality}

Let us consider $M$-theory, with its fundamental membrane and solitonic 
fivebrane, on $R^6 \times M_1 \times {\tilde M_4}$ where $M_1$ is a
one-dimensional compact space of radius $R$ and ${\tilde M_4}$ is a
four-dimensional compact space of volume $V$. We may obtain a fundamental
string on $R^6$ by wrapping the membrane around $M_1$ and reducing on
${\tilde M_4}$. Let us denote fundamental string sigma-model metrics in
$D=10$ and $D=6$ by $G_{10}$ and $G_{6}$. Then from the corresponding
Einstein Lagrangians   
\be
\sqrt{-G_{11}}R_{11}=
R^{-3}\sqrt{-G_{10}}R_{10}=\frac{V}{R}\sqrt{-G_{6}}R_{6}
\ee
we may read off the strength of the string couplings in $D=10$ \cite{DMW}
\be
{\lambda_{10}}^2=R^3
\ee
and $D=6$
\be
{\lambda_{6}}^2=\frac{R}{V}
\ee
Similarly we may obtain a solitonic string on $R^6$ by wrapping the 
fivebrane around ${\tilde M_4}$ and reducing on $M_1$. Let us denote
the solitonic string sigma-model metrics in $D=7$ and $D=6$ by ${\tilde
G}_{7}$ and ${\tilde G}_{6}$. Then from the corresponding Einstein
Lagrangians   \be 
\sqrt{-G_{11}}R_{11}=
V^{-3/2}\sqrt{-{\tilde G}_{7}}{\tilde R}_{7}=
\frac{R}{V}\sqrt{-{\tilde G}_{6}}{\tilde R}_{6} 
\ee 
we may read off the strength of the string couplings in $D=7$ \cite{DMW} 
\be
{{\tilde \lambda}_{7}}^2=V^{3/2}
\ee
and $D=6$
\be
{{\tilde \lambda}_{6}}^2=\frac{V}{R}
\ee
Thus we see that the fundamental and solitonic strings are related by a 
strong/weak coupling:
\be
{{\tilde \lambda}_{6}}^2=1/{\lambda_{6}}^2
\ee

\begin{table}
$
\begin{array}{ccccc}
~~~~~~~~~~~~~~{\bf (N_+,N_-)}&{\bf M_1}&{\bf {\tilde
M}_4}&{\bf fundamental~string}&{\bf dual~string}\\ ~~~~~~~~~~~~~~&~&~&\\
~~~~~~~~~~~~~~(1,0)&S^1/Z_2&K3&heterotic&heterotic\\
~~~~~~~~~~~~~~(1,1)&S^1&K3&Type~IIA&heterotic\\
~~~~~~~~~~~~~~(1,1)&S^1/Z_2&T^4&heterotic&Type~IIA\\
~~~~~~~~~~~~~~(2,2)&S^1&T^4&Type~IIA&Type~IIA
\end{array}
$
\label{1}
\caption{String/string dualities}
\end{table}

We shall be interested in $M_1=S^1$ (in which case from ($ii$) of section
(\ref{cool}) the fundamental string will be Type $IIA$) or $M_1=S^1/Z_2$
(in which  case from ($xiii$)  of section (\ref{cool}) the fundamental
string will be heterotic $E_8 \times E_8$). Similarly, we will be
interested in ${\tilde M}_4=T^4$ (in which case from ($ix$) of section
(\ref{cool}) the solitonic string will be Type $IIA$) or ${\tilde M}_4=K3$
(in which case from ($ix$) of section (\ref{cool}) the solitonic string
will be heterotic). Thus there are four possible scenarios which are
summarized in Table 1. $(N_+,N_-)$ denotes the $D=6$ spacetime
supersymmetries. In each case, the fundamental string will be weakly
coupled as we shrink the size of the wrapping space $M_1$ and the dual
string will be weakly coupled as we shrink the size of the wrapping space
${\tilde M}_4$. 

In fact, there is in general a topological
obstruction to wrapping the fivebrane around  ${\tilde M}_4$ provided by
(\ref{kquant}) because the fivebrane cannot wrap around a $4$-manifold that
has $n\neq 0$\footnote{Actually,
as recently shown in \cite{Wittenflux}, the object which must have integral
periods is not $T_3K_4/2\pi$ but rather $T_3K_4/2\pi - p_1/4$ where $p_1$ is the
first Pontryagin class.  This will not affect our conclusions, however.}.  This
is because the anti-self-dual $3$-form field strength $T$ on the worldvolume of
the fivebrane obeys \cite{TownsendM,Wittenfive}   \be dT=K_4 \ee
and the existence of a solution for $T$ therefore requires that $K_4$ must
be cohomologically trivial. For $M$-theory on $R^6 \times S^1/Z_2 \times
T^4$ this is no problem. For $M$ theory on $R^6 \times S^1/Z_2 \times K3$,
with instanton number $k$ in one $E_8$ and $24-k$ in the other, however,
the flux of $K_4$ over $K3$ is 
\cite{DMW} 
\be 
n=12-k
\ee
Consequently, the $M$-theoretic explanation of heterotic/heterotic duality
requires $E_8 \times E_8$ with the symmetric embedding $k=12$. This
has some far-reaching implications. For example, the duality exchanges
gauge fields that can be seen in perturbation theory with gauge fields of a
non-perturbative origin \cite{DMW}.

The dilaton $\tilde \Phi$, the string
$\sigma$-model metric $\tilde G_{MN}$ and $3$-form field strength $\tilde
H$ of the dual string are related to those of the fundamental string,  $\Phi$,
$G_{MN}$ and $H$ by the replacements \cite{Lublack,Minasian1}      
\[
\Phi \rightarrow \tilde \Phi=-\Phi
\]
\[
G_{MN} \rightarrow \tilde G_{MN}=e^{-\Phi}G_{MN}
\] 
\be
H \rightarrow \tilde H=e^{-\Phi}*H
\la{discrete}
\ee
In the case of heterotic/Type $IIA$ duality and Type $IIA$/heterotic
duality, this operation takes us from one string to the other, but
in the case of heterotic/heterotic duality and Type $IIA$/Type $IIA$
duality this operation is a discrete symmetry of the theory. This 
Type $IIA$/Type $IIA$ duality is discussed in \cite{Senvafa}
and we recognize this symmetry as subgroup of the
$SO(5,5;Z)$ $U$-duality \cite{Luduality,Hulltownsend,Dijkgraaf} of the
$D=6$ Type $IIA$ string. 

Vacua with $(N_+,N_-)=(1,0)$ in $D=6$ have been the subject
of much interest lately. In addition to DMW  vacua \cite{Minasian2}
discussed above, obtained from $M$-theory on $S^1/Z_2 \times K3$, there
are also the GP vacua \cite{Pradisi,Bianchi,Gimon} obtained from the $SO(32)$
theory on $K3$ and the MV vacua \cite{Morrison1,Morrison2} obtained from {\it
$F$-theory} \cite {Vafa} on Calabi-Yau.  Indeed, all three categories are
related by duality
\cite{Alda1,Morrison1,Gross,Wittenphase,Ferrara2,Berkooz,Morrison2,Alda2}.
In particular, the DMW heterotic strong/weak coupling duality gets mapped
to a $T$-duality of the Type $I$ version of the $SO(32)$ theory, and the
non-perturbative gauge symmetries of the DMW model arise from small
$Spin(32)/Z_2$ instantons in the heterotic version of the $SO(32)$ theory
\cite{Berkooz}.  Because heterotic/heterotic duality interchanges
worldsheet and spacetime loop expansions -- or because it acts by duality
on $H$ -- the duality exchanges the tree level Chern-Simons contributions
to the Bianchi identity \[ dH=\alpha'(2\pi)^2X_4  \] 
\be
X_4=\frac{1}{4(2\pi)^2}[\tr R^2-\Sigma_\alpha v_\alpha \tr F_\alpha{}^2] 
\la{Bianchi} 
\ee
with the  one-loop Green-Schwarz corrections to the field equations
\[
d\tilde H=\alpha'(2\pi)^2\tilde X_4
\]
\be
\tilde X_4=\frac{1}{4(2\pi)^2}[\tr R^2-\Sigma_\alpha {\tilde v}_\alpha 
\tr F_\alpha{}^2]
\la{field}
\ee
Here $F_\alpha$ is the field strength of the $\alpha^{th}$ component
of the gauge group, $\tr$ denotes the trace in the fundamental
representation, and $v_\alpha,\tilde v_\alpha$ are constants.  In fact,
the Green-Schwarz anomaly cancellation mechanism in six dimensions
requires that the anomaly eight-form $I_8$ factorize as a product of
four-forms,  
\be 
I_8=X_4\tilde X_4,
\la{eightform} 
\ee
and a six-dimensional string-string duality with the general features
summarized above would exchange the two factors \cite{Minasian1}. 
Moreover, supersymmetry relates the coefficients $v_\alpha,\tilde
v_\alpha$ to the gauge field kinetic energy.  In the Einstein  metric
$G^c{}_{MN}=e^{-\Phi/2}G{}_{MN}$, the exact dilaton dependence of the
kinetic energy of the gauge field $F_\alpha{}_{MN}$, is \cite{Sagnotti} 
\be 
L_{gauge}=-\frac{(2\pi)^3}{8\alpha'}\sqrt{G^c}\Sigma_\alpha \left( v_\alpha
e^{-\Phi/2}  +\tilde v_\alpha e^{\Phi/2} \right)\tr
F_\alpha{}_{MN}F_\alpha{}^{MN}. 
\la{happiness}
\ee
So whenever one of the $\tilde v_\alpha$ is negative, there is a value of
the dilaton for which the coupling constant of the corresponding gauge
group diverges. This is believed to signal a phase transition associated
with the appearance of tensionless strings
\cite{Ganor,Seibergwitten,Dufflupope}.  This does not happen for the
symmetric embedding discussed above since the perturbative gauge fields
have $v_\alpha>0$ and $\tilde v_\alpha=0$ and the non-perturbative gauge
fields have $v_\alpha=0$ and $\tilde v_\alpha>0$. Another kind of
heterotic/heterotic duality may arise, however, in vacua where one may
Higgs away that subset of gauge fields with negative $\tilde v_\alpha$,
and be left with gauge fields with $v_\alpha=\tilde v_\alpha>0$.  This
happens for the non-symmetric embedding $k=14$ and the appearance
of non-perturbative gauge fields is not required 
\cite{Alda1,Alda2,Johnson,Morrison1,Morrison2}.
Despite appearances, it known from $F$-theory that the $k=12$
and $k=14$ models are actually equivalent \cite{Morrison1,Morrison2}.

Vacua with $(N_+,N_-)=(2,0)$ arising from Type $IIB$ on $K3$ also have an
$M$-theoretic description, in terms of compactification on $T^5/Z_2$
\cite{Dasgupta,Wittenfive}.

\section{Four dimensions}
\label{four}

It is interesting to consider further toroidal compactification to four
dimensions, replacing $R^6$ by $R^4\times T^2$.  Starting with
a $K3$ vacuum in which the $E_8\times E_8$ gauge symmetry is completely
Higgsed, the toroidal compactification to four dimensions gives an
$N=2$ theory with the usual three vector multiplets $S$, $T$ and $U$ related
to the four-dimensional heterotic string coupling constant and
the area and shape of the $T^2$.  When reduced to four dimensions,
the six-dimensional string-string duality (\ref{discrete}) becomes
\cite{Duffstrong} an operation that exchanges $S$ and $T$, so in the case
of heterotic/heterotic duality we have a discrete $S-T$ interchange
symmetry.  This self-duality of heterotic string vacua does not rule out
the the possibility that in $D=4$ they are also dual to  Type $IIA$
strings compactified on  Calabi-Yau manifolds. In fact, as discussed in 
\cite{Kachru}, when the gauge group is completely Higgsed, obvious
candidates are provided by Calabi-Yau manifolds with hodge numbers
$h_{11}=3$ and $h_{21}=243$, since these have the same massless field
content. Moreover, these manifolds do indeed exhibit the $S-T$ interchange
symmetry \cite{Klemm,Duffelectric,Cardoso}. Since the heterotic string on
$T^2\times K3$ also has $R$ to $1/R$ symmetries that exchange $T$ and $U$,
one might expect a complete $S-T-U$ triality  symmetry, as discussed in
\cite{Duffliurahmfeld}. In all known models, however, the $T-U$
interchange symmetry is spoiled by non-perturbative effects
\cite{Louis,Cardoso1}.

An interesting aspect of the Calabi-Yau manifolds $X$ appearing in
the duality between heterotic strings on $K3 \times T^2$ and Type $IIA$
strings on $X$, is that they can always be written in the form of a $K3$
fibration \cite{Klemm}.  Once again, this ubiquity of $K3$ is presumably a
consequence of the interpretation of the heterotic string as the $K3$
wrapping of a fivebrane. Consequently, if $X$ admits {\it two}
different $K3$ fibrations, this would provide an alternative explanation
for heterotic dual pairs in four dimensions
\cite{Gross,Morrison1,Morrison2} and this is indeed the case for the
Calabi-Yau manifolds discussed above.

\section{Eleven to twelve: is it still too early?}
\label{twelve}

The $M$-theoretic origin of the Type $IIB$ string given in
$(xii)$ of section \ref{cool} seems to require going down to nine
dimensions and then back up to ten. An obvious question, therefore, is
whether Type $IIB$ admits a more direct higher-dimensional explanation,
like Type $IIA$.  Already in 1987 it was suggested
\cite{Blencowe} that the $(1,1)$-signature worldsheet of the Type $IIB$
string, moving in a $(9,1)$-signature spacetime, may be descended from
a supersymmetric extended object with a $(2,2)$-signature worldvolume,
moving in a $(10,2)$-signature spacetime. This idea becomes even more
appealing if one imagines that the $SL(2,Z)$ of the Type $IIB$ theory
\cite{Hulltownsend} might correspond to the modular group of a $T^2$
compactification from $D=12$ to $D=10$ just as the $SL(2,Z)$ of $S$-duality
corresponds to the modular group of a $T^2$ compactification from $D=6$ to
$D=4$ \cite{Duffstrong}. In view of our claims that $D=11$ is the maximum
spacetime dimension admitting a consistent supersymmetric theory,
however, this twelve dimensional idea requires some explanation. So let us
begin by recalling the $D=11$ argument.

As a $p$-brane moves through spacetime, its trajectory is described by
the functions $X^M (\xi)$ where $X^M$ are the spacetime coordinates ($M =
0, 1, \ldots, D - 1$) and $\xi^i$ are the worldvolume coordinates ($i = 0,
1, \ldots, d - 1$).  It is often convenient to make the so-called
 ``static gauge choice'' by making the $D = d + (D - d)$ split
\be
{X^M (\xi) = (X^{\mu} (\xi), Y^m (\xi)),}
\ee
where $\mu = 0, 1, \ldots, d - 1$~and~$m = d, \ldots, D - 1$,
and then setting
\be
{X^{\mu} (\xi) = \xi^{\mu}.}
\ee
Thus the only physical worldvolume degrees of freedom are given
 by the $(D -d)~Y^m (\xi)$.  So the number of on-shell bosonic degrees of
freedom is
\be
{N_B = D - d.}
\ee
To describe the super $p$-brane we augment the $D$ bosonic coordinates $X^M
(\xi)$ with anticommuting fermionic coordinates $\theta^{\alpha} (\xi)$.
Depending on $D$, this spinor could be Dirac, Weyl, Majorana or
Majorana-Weyl. The fermionic $\kappa$-symmetry means that half of the
spinor degrees of freedom are redundant and may be eliminated by a physical
gauge choice.  The net result is that the theory exhibits a {\it
$d$-dimensional worldvolume supersymmetry} \cite{Achucarro} where the
number of fermionic generators is exactly half of the generators in the
original spacetime supersymmetry.  This partial breaking of supersymmetry
is a key idea.  Let $M$ be the number of real components of the minimal
spinor and $N$ the number of supersymmetries in $D$ spacetime dimensions
and let $m$~and~$n$ be the corresponding quantities in $d$ worldvolume
dimensions.  Since $\kappa$-symmetry always halves the number of fermionic
degrees of freedom and (for $d > 2$) going on-shell halves it again, the
number of on-shell fermionic degrees of freedom is
\be
{N_F = {1\over 2}~mn = {1\over 4}~MN.}
\ee
Worldvolume supersymmetry demands $N_B = N_F$ and hence
\be
{D - d = {1\over 2}~mn = {1\over 4}~MN.}
\la{bosefermi}
\ee
We note in particular that $D_{{\rm max}} = 11$ since $M=32$ for
$D=11$ and we find the supermembrane with $d=3$. For $D \geq
12$, $M \geq 64$ and hence (\ref{bosefermi}) cannot be satisfied. 
Actually, the above argument is strictly valid only for 
$p$-branes whose worldvolume degrees of freedom are described by scalar
supermultiplets. There are also $p$-branes with vector and/or
antisymmetric tensor supermultiplets on the worldvolume
\cite{Callan1,Callan2,Luscan}, but repeating the argument still yields
$D_{{\rm max}} = 11$ where we find a superfivebrane with $d=6$
\cite{Khuristring}.   

The upper bound of $D=11$ is thus a consequence of the jump from $M=32$ to 
$M=64$ in going from  $D=11$ to $D=12$.  However, this jump can be
avoided if one is willing to pay the price of changing the signature 
to $(10,2)$ where it is possible to define a spinor which is both
Majorana and Weyl. A naive application of the above bose-fermi
matching argument then yields $D_{{\rm max}} = 12$ where we find an
extended object with $d=4$ but with $(2,2)$ signature \cite{Blencowe}.  
The chiral nature of this object then naturally suggests a connection
with the Type $IIB$ string in $D=10$, although the $T^2$
compactification would have to be of an unusual kind in order to
preserve the chirality.  Moreover, the chiral $(N_+,N_-)=(1,0)$
supersymmetry algebra in $(10,2)$ involves the anti-commutator
\cite{Bars}  
\be
\{Q_{\alpha},Q_{\beta}\}=\Gamma^{MN}{}_{\alpha \beta}P_{MN}
+\Gamma^{MNPQRS}{}_{\alpha \beta}Z^+{}_{MNPQRS}
\ee
The absence of translations casts doubt on the naive application of the
bose-fermi matching argument, and the appearance of the
self-dual $6$-form charge $Z$ is suggestive of a sixbrane, rather than
a threebrane. 

Despite all the objections one might raise to a world with
two time dimensions, and despite the above problems of interpretation, the
idea of a $(2,2)$ object moving in a $(10,2)$ spacetime has recently been
revived in the context of {\it $F$-theory} \cite{Vafa}, which
involves Type $IIB$ compactification where the axion and dilaton from
the RR sector are allowed to vary on the internal manifold.  Given a
manifold $M$ that has the structure of a fiber bundle whose fiber is
$T^2$ and whose base is some manifold $B$, then 
\be
F~ on ~M \equiv ~Type~IIB~on~B      
\ee
The utility of $F$-theory is beyond dispute and it has certainly enhanced
our understanding of string dualities, but should the twelve-dimensions
of $F$-theory be taken seriously? And if so, should  $F$-theory be
regarded as more fundamental than $M$-theory? Given that there seems to be
no supersymmetric field theory with $SO(10,2)$ Lorentz
invariance \cite{Nishino}, and given that the on-shell states carry only
ten-dimensional momenta \cite{Vafa}, the more conservative interpretation
is that the twelfth dimension is merely a mathematical artifact and that
$F$-theory should simply be interpreted as a clever way of compactifying
the $IIB$ string \cite{Sen}.  Time will tell.   

\section{Conclusion}

The overriding problem in superunification in the coming years will be to
take the Mystery out of $M$-theory, while keeping the Magic and the
Membranes. 



\end{document}